\documentclass[prd,twocolumn,tightenlines,showpacs,nofootinbib,amsfonts,amssymb,amsmath,color]{revtex4-1}
\usepackage{graphicx}
\usepackage{color}

\begin{document}
\newcommand{\etal}{{\it et al.}}
\newcommand{\bx}{{\bf x}}
\newcommand{\bn}{{\bf n}}
\newcommand{\bk}{{\bf k}}
\newcommand{\dd}{{\rm d}}
\newcommand{\dslash}{D\!\!\!\!/}
\def\ga{\mathrel{\raise.3ex\hbox{$>$\kern-.75em\lower1ex\hbox{$\sim$}}}}
\def\la{\mathrel{\raise.3ex\hbox{$<$\kern-.75em\lower1ex\hbox{$\sim$}}}}
\def\beq{\begin{equation}}
\def\eeq{\end{equation}}

\leftline{UMN--TH--3322/14,\ ACFI-T14-03}

\vskip-2cm
\title{A falsely fat curvaton with  an observable running of the spectral tilt}

\author{Marco Peloso$^1$, Lorenzo Sorbo$^{2,3}$,  
Gianmassimo Tasinato$^4$}
\affiliation{
${^1}$  School of Physics and Astronomy, University of Minnesota, Minneapolis, 55455, USA\\
${^2}$ Department of Physics, University of Massachusetts, Amherst, MA 01003, USA\\
${^3}$ Institut de Physique Th\'eorique, CEA, F-91191 Gif-sur-Yvette Cedex, France\\
${^4}$ Institute of Cosmology \& Gravitation, University of Portsmouth, Dennis Sciama Building, Portsmouth, PO1 3FX, United Kingdom\\
}
\vspace*{2cm}

\begin{abstract}

In slow roll inflation, the running of the spectral tilt is generically proportional to the square of the deviation from scale invariance, $\alpha_s \propto \left( n_s - 1 \right)^2$, and it is therefore currently undetectable. We present  a mechanism able to generate a much larger running within slow roll. The mechanism is based on a curvaton field with a large mass term, and a time evolving normalization. This may happen for instance to the angular direction of a complex field in presence of an evolving radial direction. At the price of a single tuning between the mass term and the rate of change of the normalization, the curvaton can be made effectively light at the CMB scales, giving a spectral tilt in agreement with observations. The lightness is not preserved at later times, resulting in a detectable running of the spectral tilt.  This mechanism shows that fields with a large mass term  do not necessarily decouple from the inflationary physics, and provides a new tool for model building in inflation.  

\vspace{1cm}

\end{abstract}
 \date{January 2014}
 \maketitle

\section{Introduction}
 \label{sec:intro}

Observations strongly suggest indicate that the spectrum of primordial curvature fluctuations has an almost scale invariant, nearly Gaussian distribution. The simplest set-up that explains these features is single-field, slow-roll inflation with canonical kinetic terms.  In this scenario, a light scalar field slowly rolling on a potential induces  a phase of cosmological acceleration (see~\cite{Linde:2005ht} for a review). While the simplest inflationary set-ups based on a single field provide a good fit to observations, well motivated extensions of the minimal models exist, for example including additional scalar fields that participate to the inflationary dynamics as in multiple field inflation~\cite{Wands:2007bd}, or scalars that acquire a non-trivial, interesting dynamics only after inflation ends, as in the curvaton mechanism~\cite{Linde:1996gt, Enqvist:2001zp,Lyth:2001nq,Moroi:2001ct}.

The power spectrum $P$ of the primordial perturbations that caused the observed CMB anisotropies is nearly scale invariant. Using the Taylor expansion
\begin{equation}
P \left( k \right) \equiv P \left( k_0 \right) \left( \frac{k}{k_0} \right)^{n_s - 1 + \frac{1}{2} \alpha_s \ln \frac{k}{k_0} } \;\;
\label{parametrizationP}
\end{equation}
around the pivot scale $k_0 = 0.05 {\rm Mpc}^{-1}$, the Planck mission,  supplemented with the 9-year WMAP polarization data (WP),  obtained  \cite{Ade:2013uln} 
\begin{eqnarray}
& & n_s-1  =  - 0.0397 \pm 0.009 \;\;, \nonumber\\
& & \alpha_s = - 0.013 \pm 0.009 \;\;,\;\; (68\% {\rm CL, Planck + WP}) \;\;, 
\label{ns-as-Planck}
\end{eqnarray}
The parameter $\alpha_s$ is conventionally called  the running of the spectral tilt \cite{Kosowsky:1995aa}, since  $\alpha_s \left( k \right) = \frac{d n_s \left( k \right) }{  d {\rm ln } k}$. 

Consider a conventional curvaton mechanisms, in which the isocurvature  fluctuations of a test field during inflation (namely, a field which gives a negligible contribution to the expansion) are converted into curvature pertubations after  inflation ends. A scalar massless field in de Sitter spacetime has a scale invariant power spectrum. This is not the case for a field of mass $m$ in an inflationary background, where the Hubble rate $H$ varies according to the slow roll relation  $\epsilon = - \frac{\dot{H}}{H^2} \ll 1$.~\footnote{Following the standard convention, we denote by a dot (prime)   a derivative with respect to physical (conformal) time.} Specifically, the perturbations of a scalar field $\sigma$ with canonical kinetic term obey the equation 
\begin{eqnarray}
& & \left( a\, \delta \sigma \right)'' + \left[ k^2 - \frac{2}{\tau^2} + \frac{m_{\rm eff}^2}{H^2 \tau^2} \right]  \left( a\, \delta \sigma \right) = 0 \;\;  
\label{sig-can}
\end{eqnarray}
with
\begin{eqnarray}
&& m_{\rm eff}^2 \equiv m^2 + \epsilon \left[ - 3 H^2 + 2 m^2 \left( 1 + {\rm ln } \frac{1}{-H \tau} \right) \right] + {\rm O } \left( \epsilon^2 \right) \;\;, \nonumber\\ 
\label{m-eff}
\end{eqnarray} 
leading to a spectral tilt $n_s-1 = {\rm O } \left( \frac{m_{\rm eff}^2}{H^2} \right) = {\rm O } \left( \frac{m^2}{H^2} , \epsilon \right)$.  

Therefore, in generic curvaton models, the fact that the observed cosmological perturbations  are nearly scale invariant indicates that the inflationary expansion was close to de Sitter, $\epsilon \ll 1$, and that the field was light, $m^2 \ll H^2$. The effective mass of the perturbations  varies because of the last term in the square parenthesis of (\ref{m-eff}), and because of the ${\rm O } \left( \epsilon^2 \right)$ terms,  giving rise to running of the spectral tilt.   Eq. (\ref{m-eff}) shows that  in generic curvaton models this variation is at most of ${\rm O } \left( \epsilon \frac{m^2}{H^2} , \epsilon^2 \right)$, namely 
\begin{equation}
|\alpha_s| \la {\rm O } \left( \left(  n_s -1 \right)^2 \right)  = {\rm O } \left( 10^{-3} \right) \;. 
\label{as-ns}
\end{equation} 

An analogous conclusion is reached in models of single field inflation. In this case,
\begin{equation}
n_s-1 = 2 \eta_V - 6 \epsilon_V \;\;,\;\; 
\alpha_s = - 2 \xi_V^2 + 8 \epsilon_V \left( n_s - 1 \right)  + 24 \epsilon_V^2 \;\;,
\label{singlefield}
\end{equation} 
where we introduce the slow roll parameters
\begin{equation}
\epsilon_V \equiv \frac{M_p^2}{2} \left( \frac{V_{,\phi}}{V} \right)^2 ,\; 
\eta_V \equiv M_p^2  \frac{V_{,\phi\phi}}{V}  ,\; 
\xi_V^2 \equiv M_p^4  \frac{V_{,\phi} V_{,\phi\phi\phi}}{V^2}  \;.  
\label{def-slow}
\end{equation}
(we denote with a comma the derivative of a function with respect to its argument). In generic models, the third slow roll parameter is small, and again the prediction (\ref{as-ns}) is obtained~\cite{Chung:2003iu}. 

Some past observational results have been in slight tension with this estimate.  For instance, the first WMAP release gave~\cite{Bennett:2003bz}  $\alpha_s = -0.031_{-0.018}^{+0.016}$, and, more recently, the  SPT high-$\ell$ CMB data gave~\cite{Hou:2012xq} $\alpha_s = -0.024 \pm 0.011$ (SPT + WMAP7) and $\alpha_s =  -0.028 \pm 0.010 $ (using also BAO and $H_0$ measurements). Although in agreement with the $\alpha_s = {\rm O } \left( \left(  n_s-1 \right)^2 \right)$ prediction, the Planck result~(\ref{ns-as-Planck}) is not conclusive on this regard, since it is also statistically compatible with the SPT measurement.  Therefore, if future measurements will again provide an indication for $\alpha_s = {\rm O } \left( - 0.01 \right)$ and statistically incompatible with zero, they will pose a serious challenge on standard inflationary models. See \cite{Adshead:2010mc} for a comprehensive analysis of how future large scale structure or 21 cm observations can lower current bounds on $\alpha_s$ to the region $|\alpha_s|\le 10^{-3}$, in absence of detection. In single field inflation, without  the third derivative term $\xi_V^2$, the observed spectral tilt  $n_s-1 \simeq -0.04$ is incompatible with $\alpha_s \la -0.01$.  Besides being hard to motivate, it is difficult to preserve a large third derivative, while keeping the first two  small, for a sufficiently long duration of inflation~\cite{Chung:2003iu,Easther:2006tv}, so that  the models that achieve a large running have potentials with some bump-like feature or superimposed oscillations  \cite{Hannestad:2000tj,Chung:2003iu, Feng:2003mk,Kobayashi:2010pz,Takahashi:2013tj}, or possess some peculiar aspects beyond standard scenarios~\cite{Kawasaki:2003zv,Huang:2003zp,BasteroGil:2003bv,Yamaguchi:2003fp,Ballesteros:2005eg}. 

In  curvaton models, the above estimate $\alpha_s = {\rm O } \left( \epsilon \frac{m^2}{H^2} \right)$ could be compatible with a running of first order in slow roll if $m$ was of order $H$. However, as commented after eq. (\ref{m-eff}), this would lead to an unacceptable value for $n_s$. In this work we show that $m \geq H$ is possible if the curvaton Lagrangian has  a non-standard normalization. Specifically, let us consider the case in which  the curvaton has a time-dependent normalization that we encode in an overall function that multiplies the curvaton Lagrangian: ${\cal N}^2 \left( t \right) {\cal L}_{\rm curv}$, and that a  mass term $m^2 \sigma^2$ is present \footnote{ With `mass term' of a field we  refer to the second derivative of the potential for canonically normalized fields {\em computed on a static background}. Note that this definition, including the requirement of a static background, is consistent with the standard meaning of mass. One could alternatively denote the mass from the dispersion relation of the canonically normalized variable in the $k \rightarrow 0$ limit. If we did so, we would conclude from eq. (\ref{sig-can}) that a field with $m=0$ and ${\cal N} = 1$ has the tachyonic mass  $\sim -2\,H^2$ on a (quasi)-de Sitter, non static, background. A field with $m=0$ and ${\cal N}=1$ is typically denoted as massless, and our choice of denoting as mass term the second derivative of the potential in a static background is consistent with this.} in $ {\cal L}_{\rm curv}$. This model can lead  to a  scale invariant spectrum for any value of  $m$ (hence even for $m \gg H$), provided that  ${\cal N} \left( t \right)$ has a suitable time evolution. Once we canonically normalize the curvaton field, a new term, related to the rate of change of the normalization, $\frac{\dot{\cal N}}{\cal N}$,  appears in the dispersion relation of the curvaton perturbations,  and the total $m_{\rm eff}^2$ can be tuned to a sufficiently small value by canceling this new term against those appearing in   (\ref{m-eff}). In the specific model that we will consider,  $\frac{\dot{\cal N}}{\cal N} \propto H$, and this quantity varies at first order in slow roll. On the other hand, the mass term $m$ in the curvaton Lagrangian is a constant. Therefore, the tuning that sets $m_{\rm eff}$ to the value required by $n_s - 1 \simeq 0.04$ cannot work at all scales. This naturally introduces a running of the spectral tilt which is of the first order in slow roll, and which therefore can be much larger than the prediction (\ref{as-ns}). 

\smallskip

Rather than introducing an unspecified external function of time, we can assume that ${\cal N}$ is actually a function of the inflaton,  given that the latter acts as a clock during inflation. This introduces a direct coupling between the inflaton and the curvaton perturbations.   We study the conditions for which the effect of such couplings is suppressed, so that only the inflaton zero mode dynamics affects the perturbations of the curvaton through the time evolution of ${\cal N}$. 

\medskip

From a theoretical perspective, our system shows that fields with a very large mass term do not necessarily decouple during inflation, and can play an active and distinctive role for the generation of an almost scale invariant spectrum of curvature fluctuations, if they are appropriately coupled to other degrees of freedom (as those controlling the time dependent normalization ${\cal N}$ in our example). 

This model also helps reformulating the naturalness problem in inflation. While it is usually assumed that a necessary condition for  the generation of a scale-invariant power spectrum during inflation is  the presence of a scalar field with a small mass term, in the present scenario it is possible to obtain scale invariance even for a curved potential  by appropriately tuning the coupling of the inflaton to the curvaton. Therefore the fine-tuning of the curvaton mass term to zero generalizes to a relation -- given by  eq.~(\ref{alpha-cond}) below -- involving the mass term of the curvaton and the coupling, measured by $\alpha$, to the inflaton. Note that this implies that we are {\em not} solving the so-called $\eta$-problem, as we are not aware of any symmetry that can enforce the relation~(\ref{alpha-cond}), but we are formulating it in a more general way.
 
Constructions of the form considered in this paper, where the Lagrangian of the curvaton is multiplied by a function of the inflaton, appear quite naturally in models where the inflaton is the radial part and the curvaton is the angular part of a complex field. We show this with an explicit example,   featuring a model of broken U(1) symmetry\footnote{Also in the  proposal  of  \cite{Rubakov:2009np} the curvaton is the angular field of a U(1) symmetric model, and its perturbations are controlled by the evolution of the radial direction.  In the model of \cite{Rubakov:2009np}, however, the U(1) symmetry is  unbroken, and therefore the curvaton is massless, during the epoch in which perturbations are generated.}. Moreover, by a suitable change of frame, we show that our mechanism can be re-expressed in terms of a scalar field with time dependent mass and sound speed, embedded in an appropriately chosen Friedmann-Robertson-Walker background.  

Various recent works have studied interesting scenarios in which massive fields play a role to characterize the features of primordial fluctuations produced during inflation. For example, in the gelaton scenario \cite{Tolley:2009fg}, a very massive scalar field is tightly coupled to the inflaton in such a way to reduce the inflaton effective sound speed and produce large non-Gaussianities of equilateral form in the inflaton sector.  In quasi-single field inflationary models (see e.g.~\cite{Chen:2009we,Chen:2009zp}), the interplay between the dynamics of inflaton and isocurvatons with Hubble mass can again lead to large non-Gaussian signatures in the inflaton fluctuations.  In reference~\cite{Dong:2010in} it was shown that heavy fields can adjust their value as a response to the inflaton expectation value, effectively changing the shape of the inflaton potential.  Our point of view is different, since we focus on the dynamics of the fluctuations of the isocurvature scalar field and their properties. In our  scenario, the suitable time-dependent normalization renders the isocurvature field effectively massless during inflation (although it would have a large mass in a static background).  As common in curvaton scenarios, we make the hypothesis that inflaton fluctuations have negligible amplitude and do not take part to the final curvature perturbations: hence, we will not be interested on possible non-Gaussian effects generated in the inflaton sector.

\smallskip

The paper is organized as follows. In Section \ref{sec:PS} we show how a massive field $\sigma$ with a mass term $m > H$ can obtain an almost scale invariant power spectrum if its normalization ${\cal N} \left( t \right)$ exhibits a suitable time evolution.   In Section \ref{sec:replace} we present a simple model in which ${\cal N} \left( t \right)$ is obtained through a coupling between $\sigma$ and the inflaton field $\phi$. This concrete model allow us to set some bounds on the mechanism, by requiring that the coupling {\em {(i)}} does not significantly affect the background inflaton evolution, {\em {(ii)}} does not modify the power spectrum of the primordial perturbations besides the effect related to ${\cal N} \left( t \right)$, and {\em {(iii)}} does not result in a too large non-Gaussianity.  In Section \ref{sec:run} we compute the running of the spectral tilt from this mechanism, showing that it can acquire a size much larger than in standard inflationary set-ups. In Section~\ref{sec:realization} we present the example of a possible realization of this mechanism in an inflationary set-up, and   in Section~\ref{sub:frame}  we discuss how our system is equivalent to that of a field in a Universe with an arbitrary expansion law, provided it has an appropriate time evolution of its mass and its speed of sound. In Section~\ref{sec:conclusions} we present our Conclusions. 

\section{A  field with a large mass term but scale invariant fluctuations }
 \label{sec:PS}

Let us consider a real scalar field $\sigma$ with a mass term $m \geq H$, where $H$ is the Hubble rate during inflation~\footnote{Our results are valid also if the mass term $m$ is smaller than the Hubble rate $H$ during inflation. However, taking $m \ll H$ reproduces the standard case of a light field with very nearly constant normalization, and therefore we target the discussion to the $m \geq H$ case.}.  For the sake of explaining our arguments in the simplest possible terms, we take $H$ as constant in the computations of this Section, assuming that the slow roll parameter $\epsilon\equiv-\dot{H}/H^2$ is negligible. The impact of a sizable  $\epsilon  $ on this mechanism will be studied in Section \ref{sec:run}. 

The action for $\sigma$ is 
\begin{equation}
S_\sigma = \int d^4 x \sqrt{-g} \;  {\cal N}^2 \left( t \right) \left( - \frac{1}{2} g^{\mu \nu} \partial_\mu \sigma \partial_\nu \sigma - \frac{m^2}{2}  \sigma^2 \right) \;\;,
\label{action1}
\end{equation}
where the only difference from a theory of a free massive field is in the assumption  that the Lagrangian of $\sigma$ is multiplied by a time-dependent function ${\cal N}$. 

First of all, we show that a suitable time dependence of ${\cal N}$ results in a scale invariant spectrum for the perturbations of $\sigma$. Specifically, we assume that ${\cal N}$ scales as a power law of the scale factor $a$ for at least  the last  $\simeq 60 $ e-folds of inflation:
\begin{equation}
{\cal N} = \left\{ \begin{array}{c} a^\alpha \;\;,\;\; a \leq 1
\\  1 \;\;,\;\; a \geq 1 \end{array} \right.  \;\;, 
\label{N-t}
\end{equation} 
(with constant $\alpha$) where the scale factor has been normalized such  that $a=1$ at the end of inflation. It is natural to try to ascribe the time dependence of (\ref{N-t}) to the  evolution of the inflaton, and we will do so in Section~\ref{sec:replace}. For the computations of this Section we can simply consider ${\cal N}$ as an external function of time. 

We use conformal time $\tau$, defined from the line element $ds^2 = a^2 \left( \tau \right) \left[ - d \tau^2 + d \vec{x}^2 \right]$, and we decompose 
\begin{equation}
\sigma(\tau, \vec{x}) \equiv \frac{ \chi(\tau, \vec{x}) }{  a \,  {\cal N} }  \equiv \frac{ 
 \chi^{(0)} \left( \tau \right) + \delta \chi \left( \tau , \vec{x} \right)  }{  a \, {\cal N} }  \;\;.
\label{phi-chi}
\end{equation}
During inflation, the Fourier transform of the perturbations of $\chi$ obeys  the equation~\footnote{Notation: For any quantum field $X$ in real space, we denote by ${\hat X}_k$ its Fourier transform, and by $X_k$ the mode functions of the Fourier transform, see also eqs.~(\ref{dc-sol}) and~(\ref{chi-quant}).} 
\begin{equation} 
\delta \chi_k'' + \left[ k^2 + \frac{m^2/H^2 - \left( \alpha+1 \right) \left( \alpha + 2 \right)}{\tau^2} \right] \delta \chi_k = 0 \;\;, 
\label{eom-dc}
\end{equation}
where we have used the fact that $ a = -{1}/{(H \tau)}$, if the variation of $H$ is negligible. See Section~\ref{sec:run} for the extension of this computation at first order in slow roll. 
  The field $\chi$ is canonically normalized in (\ref{action1}):  imposing that the mode field $\sigma$ is in the adiabatic vacuum during the sub-horizon regime  results in the early time solution $\delta \chi_{\rm in} \simeq {\rm e}^{-i k \tau} / \sqrt{2 k}$. It is then straightforward to see that the perturbations of $\sigma$ are scale-invariant in the super-horizon regime provided that 
\begin{equation}
m^2 = \alpha \left( \alpha + 3 \right) H^2 \;\;\Rightarrow\;\; \alpha = \frac{3}{2} \left( - 1 \pm \sqrt{1+\frac{4 m^2}{9 H^2}} \right) \;\;.
\label{alpha-cond}
\end{equation}
Namely, a scale invariant spectrum can be obtained for an arbitrarily large mass, provided that the normalization ${\cal N}$ varies sufficiently fast;  we find two branches of solutions, labeled with $\pm$ in eq. (\ref{alpha-cond}). In a sense,  an appropriate choice of ${\cal N}$ renders the field effectively  massless. When eq. (\ref{alpha-cond}) holds, during inflation the mode functions of $\delta \chi$ have the same solution as those of a massless scalar field with constant normalization 
\begin{equation}
\delta \chi_k = \frac{{\rm e}^{-i k \tau}}{\sqrt{2 k}} \left( i + \frac{1}{k \tau} \right) \;\;. 
\label{dc-sol}
\end{equation} 

If the field $\sigma$ is responsible for the cosmological perturbations, its power cannot be precisely scale invariant, but it needs to agree with (\ref{ns-as-Planck}). The condition (\ref{alpha-cond}) needs therefore  to be replaced by
\begin{equation} 
\alpha = \frac{3}{2} \left( - 1 \pm \sqrt{1+\frac{4 m^2}{9 H^2}} \right) \mp \frac{n_s-1}{2 \sqrt{1+\frac{4 m^2}{9 H^2}}} + {\rm O } \left( n_s - 1 \right)^2 \;\;. 
\label{alpha-ns}
\end{equation} 
As we will discuss in Section \ref{sec:run}, it is straightforward to extend this relation to the case of a slowly varying $H$, following the standard computations done for $\alpha=0$ (see for instance \cite{Riotto:2002yw}). We will learn that this leads to an interesting footprint for our scenario, namely a large
running of the spectral tilt. 

Equation (\ref{alpha-ns}) shows the change in the spectral tilt resulting from a change in $\alpha$, and therefore indicates the width of the allowed interval for $\alpha$ compatible with the experimentally allowed range $\Delta n_s$. For large field mass we obtain
\begin{equation}
\left\vert \frac{\Delta \alpha}{\alpha} \right\vert \simeq \frac{3 H^2}{4 m^2} \, \Delta n_s \;\;,\;\; m \gg H \;\;. 
\end{equation}
This equation quantifies the fine-tuning in the mechanism, namely the accuracy to which $\alpha$ needs to be set to a given value to be compatible with data at a given scale. We see that the degree of fine tuning increases as $m$ increases. The problem of preserving this tuning against radiative corrections is the manifestation in our mechanism of the so called $\eta$ problem of inflationary cosmology. 

In the remainder of this Section, to keep the algebra simple, we restrict our  discussion to the scale invariant case (\ref{alpha-cond}), but all our results  can be straightforwardly extended to the case (\ref{alpha-ns}). 

During inflation, the background field $\chi^{(0)}$  obeys an equation identical to (\ref{eom-dc}) with $k =0$. This results into 
\begin{equation}
\chi^{(0)} \left( \tau \right) = C \, a \left( \tau \right)  \;\;, 
\label{chi-0}
\end{equation} 
where $C$ in an integration constant, and where we have disregarded the rapidly decaying solution $\chi^{(0)} \propto a^{-2}$. For definiteness, we assume $C>0$. In principle, the value for $C$ could be determined only with the knowledge of the history of the universe before the final $60$ e-folds of inflation. We do not commit ourselves to any specific scenario for this previous stage, but we rather study under which conditions on $C$ our mechanism can work. 

The energy density associated with (\ref{chi-0}) is 
\begin{equation}
\rho^{(0)} = - T_0^{\,\,0} \vert_{\rm background } = \frac{\alpha \left( 2 \alpha + 3 \right) }{2} \, C^2 \, H^2 \,\,, 
\label{rho-0}
\end{equation} 
where $T_{\mu}^{\,\,\nu}$ is the energy-momentum tensor associated with (\ref{action1}).  Namely, the specific evolution of ${\cal N}$ that provides a scale invariant spectrum of $\delta \chi$ also results in a constant background energy density for $\chi$. By keeping $C$ sufficiently small, we can impose that $\rho^{(0)} \ll 3 H^2 M_p^2$, so that this energy density is negligible during inflation. 

We have also to make sure that the energy in the perturbations does not exceed the energy in the background. To do so, we expand $-T^{\,\,0}_0$ to second order, dubbing the result as $\rho^{(2)}$, and we quantize $\delta \chi$ according to 
\begin{equation}
\delta \chi = \int \frac{d^3 k}{\left( 2 \pi \right)^{3/2} } {\rm e}^{i \vec{x} \cdot \vec{k}} \, \delta {\hat \chi}_k \;\;,\;\;
\delta {\hat \chi}_k = \delta \chi_k \left( \tau \right) {\hat a}_{\vec{k}} +  \delta \chi_k^* \left( \tau \right) {\hat a}_{-\vec{k}}^\dagger  \;\;, 
\label{chi-quant}
 \end{equation} 
where the mode functions in the last expression are given by (\ref{dc-sol}), and where the operators satisfy the algebra 
$\left[ {\hat a}_{\vec{k}} ,  {\hat a}_{\vec{p}}^\dagger \right] = \delta^{(3)} \left( \vec{k} - \vec{p} \right)$. We then obtain
the theoretical expectation value 
\begin{equation}
\frac{d \,\langle \rho^{(2)} \rangle}{d \,{\rm ln k } } = \pi \alpha \left( 2 \alpha + 3 \right) \, H^4 \,\,, \;\;  \vert - k \tau \vert \ll 1 \;\; .  
\label{rho2}
\end{equation} 
The theoretical expectation value for the energy density is obtained by integrating this quantity over the modes that have left the horizon during inflation and have become classical. The integral is simple, as (\ref{rho2}) is $k-$independent, and we simply obtain  $\langle \rho^{(2)} \rangle = \pi \alpha \left(2 \alpha + 3 \right) H^4 \, N_{e}$, where $N_e$ is the number of e-folds of inflation. This quantity is typically much smaller than the inflaton energy density $3 H^2 M_p^2$. 

\smallskip

The observable we eventually care about is the curvature perturbation after inflation, under the assumption that $\sigma$ is the curvaton field responsible for it. It is conventional to compute the quantity $\zeta$, which represents the curvature perturbation on uniform density hypersurfaces $\zeta$. In spatially flat gauge ($\delta g_{ij,{\rm scalar = 0}}$, where $g$ is the metric), each species $i$ induces the curvature perturbation $\zeta_i = - H\,{ \delta \rho_i}/{\dot{\rho}_i^{(0)}}$. 
The total curvature perturbation after inflation is then given by 
\begin{equation}
\zeta = f \, \zeta_\sigma + \left( 1 - f \right) \zeta_{\rm rad} \;\;,\;\; f \equiv \frac{3 \rho_\sigma^{(0)}}{3 \rho_\sigma^{(0)} + 4 \rho_{\rm rad}^{(0)}} \;\;, 
\label{zeta-tot}
\end{equation}
where `rad' denotes the contribution from the thermal bath formed when the inflaton decays, which we are  assuming to take place well before the decay of the curvaton $\sigma$. Eq. (\ref{zeta-tot}) is actually valid after the decay of the inflaton and before that of $\sigma$. The ratio $f$ is typically very small at the end of inflation, but it then becomes of order one provided the field $\sigma$ is sufficiently long-lived. The perturbations in $\sigma$ are given by 
\begin{equation}
\zeta_\sigma \Big\vert_+ = - \frac{H \delta \rho_\sigma}{\dot{\rho}^{(0)}_\sigma} \Big\vert_+ =  \frac{ \delta \rho_\sigma}{3 \rho^{(0)}_\sigma}  \Big\vert_+  =  \frac{ \delta \rho_\sigma}{3 \rho^{(0)}_\sigma}  \Big\vert_- \,\,,  
\label{zeta-phi}
\end{equation}
where the suffix $-$ ($+$) indicates that the quantity is evaluated on super-horizon scales during (after) inflation. We note that the ratio is constant both during and after inflation. After inflation, $\sigma$ is a standard massive scalar field, and both $\rho_\sigma^{(0)} , \delta \rho_\sigma \propto a^{-3}$.~\footnote{Notice that, being very massive, the curvaton starts to oscillate around the minimum of its potential soon after inflation ends.} During inflation instead both these quantities are constant. The final equality in (\ref{zeta-phi}) is due to the fact that the detailed physics responsible for the transition in ${\cal N}$ does not affect the perturbation at super-horizon scales. It is therefore convenient to evaluate $\zeta_\sigma$ on super-horizon scales during inflation. 

Given that the energy of a massive field is $\rho_\sigma = m^2 \sigma^2$, the  ${\cal O } \left( \delta \sigma^2 \right)$ contribution to the energy density induces a non-Gaussianity  in $\zeta$, even if $\delta \sigma$ is by itself a Gaussian field. This non-Gaussianity is of the local shape, with the nonlinear parameter \cite{Lyth:2002my}:
\begin{eqnarray}
f_{\rm NL}  =   \frac{5}{4 \; r} \;\;\;\; \left( \sigma \; {\rm is \; Gaussian} \;,\; \zeta_\phi \; {\rm is \; negligible} \right)  
\label{fNL-curva}
\end{eqnarray} 
where $r$ is the value of the ratio $f$ evaluated at the curvaton decay. The Planck limit on local non-Gaussianity~\cite{Ade:2013ydc} translates into $r \ga 0.087$ at $95\%$C.L.. The contribution~(\ref{fNL-curva}) adds up to that induced by the `intrinsic' non-Gaussianity of the field $\sigma$. In Section \ref{sec:replace} we will quantify this second contribution. 

Expanding $-T^0_0$ to first order, and using~(\ref{phi-chi}), (\ref{chi-0}), and~(\ref{zeta-phi}), we obtain (disregarding slow roll corrections) 
\begin{equation}
\zeta_\sigma = - \frac{2 H \tau \left[ 2 \left(   \alpha + 2  \right) \delta \chi + \tau \delta\chi' \right] }{ 3 \left( 2 \alpha + 3 \right) C } \simeq \frac{2}{3} \, \frac{\delta \chi}{\chi^{(0)}}  \,\,, 
\label{zetasig}
\end{equation} 
where in the last expression we have used the super-horizon limit of eq. (\ref{dc-sol}). 

From this relation, we obtain the power spectrum for $\zeta_\sigma$ at super-horizon scales 
\begin{equation}
P_{\zeta_\sigma} = \frac{k^3}{2 \pi^2} \vert  \zeta_{\sigma k} \vert^2 = \frac{H^2}{9 \pi^2 C^2} \;\;, 
\label{P-zeta}
\end{equation} 
which is indeed constant. We assume that  $\sigma$ generates the observed primordial perturbation, namely that  $ r \, \zeta_\sigma > \left( 1 - r \right) \zeta_{\rm rad}$ (cf.  eq. (\ref{zeta-tot})).  We recall that  $\zeta_{\rm rad}$ is the  perturbation in the thermal bath formed at the decay of the inflaton field $\phi$. Therefore, we can identify it with $\zeta_\phi $, and assume the standard slow roll result for the latter. We therefore require that 
\begin{equation}
\gamma \equiv \sqrt{\frac{P_{\zeta_\phi}}{P_{\zeta_\sigma}} } = \frac{3 C}{2 \sqrt{2 \, \epsilon} M_p} \ll \frac{r}{1-r} = {\rm O } \left( 1 \right) \;\;, 
\label{gamma}
\end{equation}
(where we have assumed that both $r$ and $1-r$ are of order one). If this is the case, then $\zeta \simeq r\,\zeta_\sigma$, and we can identify (\ref{P-zeta}) with the observed power spectrum, which is subject to the COBE normalization $P_\zeta \simeq \left( 5 \cdot 10^{-5} \right)^2$. This gives 
\begin{equation}
\frac{H \, r}{C} \simeq \frac{3 \pi}{2} \times 10^{-4} \;\;\;,\;\;\; {\rm COBE} \;. 
\label{cobe}
\end{equation} 

In concluding this Section,  it is worth noting that  despite the fact that ${\cal N} \left( t \right)$ changes by many orders of magnitude during inflation, all physical quantities like (\ref{rho-0}), (\ref{rho2}), and (\ref{P-zeta}) are (nearly) constant during inflation. This has been achieved `by construction', as the time dependence of ${\cal N} \left( t \right)$ balances against the effect of the heavy mass $m > H$. Nonetheless it is remarkable that no pathologies arise in (\ref{action1}) and that the theory remains under perturbative control. We note that, for any given value of $m$, the required condition (\ref{alpha-cond}) is solved by both positive and negative $\alpha$. Positive $\alpha$ correspond to ${\cal N} \ll 1$ during inflation, while negative $\alpha$ correspond to  ${\cal N} \gg 1$ during inflation. As long as the action for $\sigma$ coincides with  (\ref{action1}),  both possibilities are acceptable. However, if the action contains some higher order $\sigma-$dependent terms that are not multiplied by appropriate powers of ${\cal N}$, then if ${\cal N} \ll 1$ these terms would lead to unacceptably strong couplings during inflation. This is a standard strong coupling problem, which would, at the very least, drive the theory out of perturbative control. If this is the case, we restrict our attention to the $\alpha < 0$ case, for which such additional terms are negligible during inflation. 

\section{Replacing ${\cal N} \left( t \right)$ with ${\cal N} \left( \phi \right)$}
 \label{sec:replace}

In the previous Section we have treated ${\cal N}$ as a classical external function. In any sensible model, the value of ${\cal N}$ needs to be related to the expectation value of a quantum field $\phi$. To keep the model minimal, we assume that the field $\phi$ is the inflaton. This  allows us to obtain some model specific constraints on our mechanism. The action  (\ref{action1}) is therefore extended to 
\begin{eqnarray}
S & = &   \int d^4 x \sqrt{-g} \Bigg\{ - \frac{1}{2} \left( \partial \phi \right)^2 - V \left( \phi \right) \nonumber\\ 
& & \quad\quad\quad\quad  + {\cal N}^2 \left( \phi \right) \left[ - \frac{1}{2} \left( \partial \sigma \right)^2 - \frac{m^2}{2} \sigma^2 \right] \Bigg\} \;\;. 
\label{action2}
\end{eqnarray} 
The functional form of ${\cal N}$ can be related to that of the inflaton potential by reverse-engeneering 
\cite{Ratra:1991bn,Martin:2007ue}:
\begin{equation}
{\cal N} \left( \phi \right) = {\cal N}_0 \, {\rm exp } \left[ - \int^\phi \frac{\alpha \; d \phi'}{\sqrt{2 \epsilon_V \left( \phi' \right) } \, M_p} \right] \;\; \Rightarrow \;\; \langle {\cal N \rangle} \propto a^\alpha \;\;, 
\label{N-phi}
\end{equation}
To verify the scaling, let us differentiate (\ref{N-phi}) with respect to time. We obtain the exact relation 
\begin{equation}
\langle {\cal \dot{N}} \rangle =  - \frac{\alpha}{\sqrt{2 \epsilon_V} M_p} \langle {\cal N} \rangle \dot{\phi} \;. 
\end{equation}
We can then make use of the slow roll relation \footnote{Notation: we write  ${\rm O } \left( \epsilon^n \right)$ to indicate that a quantity is of $n-$th order in any of the  slow roll parameters.} 
\begin{equation}
\dot{\phi} = - \sqrt{2 \epsilon_V} H M_p  \left[ 1 - \frac{2}{3} \epsilon_V + \frac{1}{3} \eta_V  + {\rm O } \left( \epsilon^2 \right) \right] \;, 
\end{equation}
and obtain (for simplicity, we focus on the case of constant $\alpha$)
\begin{eqnarray}
\langle {\cal \dot{N}} \rangle & = &    \alpha \langle {\cal N} \rangle  H   \left[ 1 - \frac{2}{3} \epsilon_V + \frac{1}{3} \eta_V  + {\rm O } \left( \epsilon^2 \right) \right] \;,  \nonumber\\ 
\langle {\cal \ddot{N}} \rangle & = &    \alpha^2 \langle {\cal N} \rangle  H^2   \left[ 1 - \frac{4 \alpha+3}{3 \alpha} \epsilon_V + \frac{2}{3} \eta_V  + {\rm O } \left( \epsilon^2 \right) \right]  \;. \nonumber\\ 
\label{dotN-ddotN}
\end{eqnarray} 
We give these relations at first nontrivial order in slow roll, as this will be needed for the computations of Section \ref{sec:run}. For the present considerations, we immediately see that the leading term in $\langle {\cal \dot{N}} \rangle $ gives the required  scaling  $\langle {\cal N} \rangle \propto  a^\alpha$.
 
Let us now move to discuss some physical consequences of the couplings between inflaton and curvaton. At the background level, the  $00$ Einstein equation and the inflaton equation of motion in the model read, respectively, 
\begin{eqnarray} 
& & H^2 - \frac{1}{3 M_p^2} \left( \frac{\dot{\phi}^2}{2} + V \right) = 
\frac{{\cal N}^2}{6 M_p^2} \left( \dot{\sigma}^2 + m^2 \sigma^2 \right) \;\;, \nonumber\\ 
& & \ddot{\phi} + 3 H \dot{\phi}  + V_{,\phi} = {\cal N} \, {\cal N}_{,\phi} \left( \dot{\sigma}^2 - m^2 \sigma^2 \right) \;\;. 
\label{eom-bck} 
\end{eqnarray} 
In the computations of the previous Section we have assumed that the curvaton contribution to the background evolution is negligible. This corresponds to imposing that the right hand side (RHS) of these equations is negligible. We impose that the RHS is much smaller than the dominant contributions on the left hand side (specifically, $H^2$ in the first equation, and $3 H \dot{\phi}$ in the second one). Using eqs. (\ref{phi-chi}), (\ref{alpha-cond}), (\ref{chi-0}), and (\ref{N-phi}), we obtain, respectively,  
\begin{equation}
\frac{\alpha \left( 2 \alpha + 3 \right) C^2}{6 M_p^2} \ll 1 \;\;,\;\; 
\frac{\alpha^2 C^2}{2 M_p^2 \epsilon} \ll 1 \;\;. 
\label{backreaction}
\end{equation}
The first condition corresponds to imposing that the energy density of $\sigma$ is much smaller than that of the inflaton during inflation, and we have already commented on this after eq. (\ref{rho-0}). The second condition instead arises from the specific mechanism that we are now considering, and it amounts  in  requiring that the motion of the inflaton is not modified by its coupling to $\sigma$. As $\epsilon \ll 1$, this second condition  dominates over the first one. Loosely speaking, this indicates that it is easier for the coupling to modify the dynamics of the inflaton than that of the scale factor, since the former is  slow roll suppressed with respect to the timescale $H$ of the latter (in other words, the flatter the inflaton potential is, the more one should be concerned that the couplings of the inflaton to other fields affect the inflaton motion). Interestingly, we can rewrite this condition as (recall the definition of $\gamma$ in eq. (\ref{gamma}))
\begin{equation}
\gamma \ll \frac{3}{2 \alpha} \;\;, 
\label{bacreaction-gamma}
\end{equation}
which is more stringent than (\ref{gamma}) for $\alpha \gg 1 $. 

Let us now study the perturbations in the model. In the previous Section, ${\cal N}$ was treated as an external function of time, and it  only  affected  the curvature perturbations through the linearized equation  (\ref{eom-dc}). Since ${\cal N}$ is now  function of the inflaton, the perturbations $\delta \phi$ must also  be taken into account. We keep the same working hypothesis that we had in the previous Section, namely we impose that  the observed curvature perturbation is only due to that of the curvaton, $\zeta \simeq r \, \zeta_\sigma$. However, the inflaton perturbation $\delta \phi$ modifies the final value of $\zeta_\sigma$  through its coupling with $\delta \sigma$ that is encoded in  (\ref{action2}). 

\begin{figure}[ht!]
\centerline{
\includegraphics[width=0.4\textwidth,angle=0]{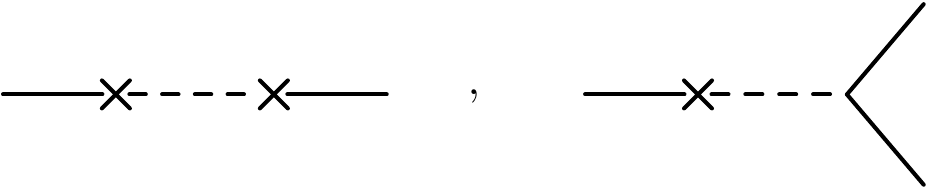}
}
\caption{Diagrams for the dominant modifications of $\langle \sigma^2 \rangle $ and $\langle \sigma^3\rangle$ due to the interaction with the inflaton $\phi$. The external solid lines refer to the curvaton $\sigma$, while the dashed lines refer to  $\phi$. The crosses denote the $\phi-\sigma$ mass insertion, proportional to the curvaton vev $C$. 
}
\label{fig:diagrams}
\end{figure}

The interactions are readily obtained by expanding ${\cal N} \left( \phi^{(0)} + \delta \phi \right)$ in the action (\ref{action2}). The dominant corrections to $\sigma^2$ and $\sigma^3$ are given by the diagrams shown in Figure \ref{fig:diagrams}. Notice in particular the interaction corresponding to a mass insertion, that plays an important role in computing correlation functions. We perform the full calculation in Appendix \ref{app:s2s3}. We obtain 
\begin{eqnarray}
& & \left\langle {\hat \zeta}_{\sigma \vec{k}_1}  {\hat \zeta}_{\sigma \vec{k}_2} \right\rangle \simeq 
\frac{2 H^2}{9 C^2} \, \left[ 1 - \frac{32 \alpha^4 \gamma^2 N_{\rm CMB}}{27} \right] 
 \frac{\delta^{(3)} \left( \vec{k}_1 + \vec{k_2} \right) }{ k_1^3 } \;\;,  \nonumber\\ 
& & \left\langle {\hat \zeta}_{\sigma \vec{k}_1}  {\hat \zeta}_{\sigma \vec{k}_2}  {\hat \zeta}_{\sigma \vec{k}_3}  \right\rangle \simeq 
- \frac{H^4 \alpha^4 N_{\rm CMB} }{12 \sqrt{2} \pi^{3/2} \epsilon^2 M_p^4 \gamma^2} \nonumber\\
& & \quad\quad \quad\quad\quad\quad  \quad\quad\quad\quad 
\times  \frac{\sum_i k_i^3}{\prod_i k_i^3} \delta^{(3)} \left( \vec{k}_1 + \vec{k}_2 + \vec{k}_3 \right) \;\;,   \nonumber\\ 
\label{s2s3-res}
\end{eqnarray} 
where the first term in the two-point function is the free field result. In these relations, $N_{\rm CMB}$ is the number of e-folds before the end of inflation when the external modes left the horizon.  As the $N_{\rm CMB}$ dependence indicates, the inflaton perturbations continue to source the perturbations of $\sigma$ all throughout the super-horizon regime. A growth with $N_{\rm CMB}$ is also obtained in the model studied in  \cite{Bartolo:2012sd}, where an analogous interaction is present between the inflaton and a vector field. 

For perturbation theory to be under control, the second term in $\langle \zeta^2 \rangle$ needs to be subdominant with respect to the free term. This gives a further condition on $\gamma$: 
\begin{equation}
\gamma \ll \sqrt{\frac{3}{2}} \, \frac{3}{4} \, \frac{1}{\alpha^2 \sqrt{N_{\rm CMB}}} \;\;, 
\label{Ds2-gamma}
\end{equation}
which is more stringent than (\ref{gamma}) and (\ref{bacreaction-gamma}). To summarize, the inflaton-curvaton interaction  associated with the mass insertion provides a correction  to the two-point function depending on $\alpha$ (that is, on the ratio $m/H$, see eq.~(\ref{alpha-cond})) and the total number of e-foldings. In order to make this correction negligible, we find the  bound (\ref{Ds2-gamma}) on  the ratio between inflaton and curvaton contributions to the curvature power spectrum. 

We can now estimate the amount of non-Gaussianity produced by the inflaton-curvaton interaction, associated with the second diagram in Fig.~(\ref{fig:diagrams}). At face value, the bispectrum given in (\ref{s2s3-res}) is of the exact local shape. However, in both expression in   (\ref{s2s3-res}) we have neglected terms that have the same parametric dependence as the terms that we have reported, but that are not enhanced by a factor $N_{\rm CMB}$. For brevity, we denote these terms as `order one contributions' in this discussion (as opposed to the ${\rm O } \left( N_{\rm CMB} \right)$ terms that we have written). In the three point function of (\ref{s2s3-res}) we have  disregarded the difference between $N_{\rm CMB}$ associated to different modes, as this difference is also an order one factor (namely, ${\rm ln }\left(\frac{1}{-k_1 \tau} \right) = {\rm ln } \left(\frac{1}{-k_2 \tau} \right)+ {\rm O } \left( 1 \right)$). For this reason,  the precise shape of the bispectrum cannot be obtained within our approximation. However, whatever the precise shape is, it is a smooth function of the momenta enhanced in the squeezed limit as the local template. Therefore, as a rough indication, we assume that the bispectrum in (\ref{s2s3-res}) is exact -- and hence it is of the local shape -- and we compare it with the bounds on local non-Gaussianity from Planck \cite{Ade:2013ydc}.

It is standard to parametrize the local non-Gaussianity by the nonlinear parameter $f_{\rm NL}$, related to the bispectrum by 
\begin{equation}
\left\langle {\hat \zeta}_{\vec{k}_2}  {\hat \zeta}_{\vec{k}_2}  {\hat \zeta}_{\vec{k}_3}  \right\rangle 
\equiv \frac{3}{10} \left( 2 \pi \right)^{5/2} f_{\rm NL} P_\zeta^2 \, \frac{\sum_i k_i^3}{\prod_i k_i^3} \delta^{(3)} \left( \vec{k}_1 + \vec{k}_2 + \vec{k}_3 \right) \,, 
\label{fNL}
\end{equation}
(the  numerical factor in this relation depends on the  $2 \pi$ convention in (\ref{chi-quant}). See 
\cite{Barnaby:2011vw} for the details). 

Under the assumption that $\zeta = r \, \zeta_\sigma$, we obtain 
\begin{equation} 
- f_{\rm NL} \simeq  \frac{20 \alpha^4 \gamma^2  N_{\rm CMB}}{9 r } \ll  \frac{15}{8 r} \,\,, 
\end{equation} 
where  the inequality follows from  the bound (\ref{Ds2-gamma}). We see that this contribution is less important than (\ref{fNL-curva}).  Perhaps not surprisingly, the ratio between the contribution to $f_{\rm NL}$ from the inflaton interaction and that of the free theory (namely, eq.  (\ref{fNL-curva})) is parametrically the same as the ratio between the contribution to the power spectrum  from the inflaton interaction and that of the free theory.    

To conclude, in this Section we have presented a more complete version of the basic mechanism introduced in Section \ref{sec:PS}, by identifying ${\cal N}$ with a function of the inflaton field. This introduces a direct coupling between the curvaton and the inflaton. In this Section we have shown that this interaction does not significantly  modify the two- and three-point correlation function of the curvaton (and, presumably, also the higher point functions: see Appendix) provided that the condition (\ref{Ds2-gamma}) is satisfied. If this is the case, the main phenomenological results (\ref{fNL-curva}) and (\ref{P-zeta}) of the previous Section continue to hold. In   Section  \ref{sec:realization} we   further characterize the mechanism by discussing how such an interaction can arise in some concrete models. 

\section{An observable running of the spectral tilt}
 \label{sec:run}

The curvaton mode functions satisfy 
\begin{eqnarray}
& & \delta \chi_k'' + \left[ k^2 + a^2 H^2 {\cal M}^2 \right]  \delta \chi_k = 0 \;, \nonumber\\ 
& &  {\cal M}^2 =  \frac{m^2}{H^2} - 2  - \frac{\dot{H}}{H^2}  - 3  \frac{\cal \dot{N}}{H{\cal N}} -  \frac{\cal \ddot{N}}{H^2{\cal N}}     \;.  
\label{dc-eom}
\end{eqnarray} 
This equation reduces to (\ref{eom-dc}) at leading order in slow roll, namely for  ${\cal N} \propto a^\alpha$ and $\dot{H} = 0$. We recall that the solutions of  (\ref{eom-dc})  exhibit a constant spectral tilt, see eq. (\ref{alpha-ns}). 

We now show that a running of the spectral tilt appears at first order in slow roll.   Specifically, we expand eq. (\ref{dc-eom})  to first order in slow roll, under the assumptions that the two slow roll parameters $\epsilon$ and $\eta$ are of comparable size. We also extend the parametrization of the mass term to  first order in slow roll:  
\begin{equation} 
m^2 \equiv \left[ \alpha \left( 3 + \alpha \right) + \delta_m \epsilon \right] H_{\rm end}^2 \,, 
\label{m-alpha-eps} 
\end{equation} 
where $H_{\rm end}$ is the value of the Hubble rate at the end of inflation, and where $\delta_m$ is a constant that we will set below to reproduce the observed spectral tilt at the CMB scales. For simplicity, we consider the parameter $\alpha$  as constant (see however the discussion at the end of this section).  We note that  $\epsilon = \epsilon_V = {\rm constant}$ to leading  order in slow roll.  

Using expressions (\ref{dotN-ddotN}), (\ref{m-alpha-eps}), and the background solutions 
\begin{eqnarray}
& & a^2 H^2  =   \frac{1+2 \epsilon}{\tau^2} + {\rm O } \left( \epsilon^2 \right) \;\;,\;\; \nonumber\\
& & H  =   H_{\rm end} \left[ 1 - \epsilon \; {\rm ln } \left( - \frac{1}{H_{\rm end} \tau} \right) \right] + {\rm O } \left( \epsilon^2 \right) 
\end{eqnarray} 
the equation   (\ref{dc-eom}) acquires the form 

\begin{eqnarray} 
&&\frac{\partial^2}{\partial_x^2} \delta \chi_k  + \left( 1 - \frac{2}{x^2} \right) \delta \chi_k 
= \epsilon \, F \left( x \right) \delta \chi_k  + {\rm O } \left( \epsilon^2 \right)   \;\;,   \nonumber\\
&& F \left( x \right) = \frac{F_0 - 2 \alpha \left( \alpha + 3 \right) \ln \frac{k}{H x}}{x^2} \;, \nonumber\\ 
&& F_0 \equiv 3-3\alpha - \frac{4}{3} \alpha^2 - \delta_m + \alpha \left( 1 + \frac{2 \alpha}{3} \right) \frac{\eta_V}{\epsilon} \;, 
\label{eom-dc-eps}  
\end{eqnarray} 
where $x \equiv - k \tau$.~\footnote{Due to the fact that we are working at first order in slow roll, we can disregard the variation of $H$  in the expression for $F_0$ in eq.~(\ref{eom-dc-eps}), and in the equations below.} 

Notice that eq. (\ref{eom-dc-eps}) at zeroth order in slow roll reduces to  the equation for the modes of a massless field in de Sitter background. This is because the parametrization (\ref{m-alpha-eps}) has been chosen to satisfy the condition  (\ref{alpha-cond}) at zeroth order. We recall that the condition  (\ref{alpha-cond}) expresses the fine-tuning necessary for a scale invariant spectrum in de Sitter background. This relation was  replaced by (\ref{alpha-ns}), to allow for a departure from exact scale invariance. In eq. (\ref{m-alpha-eps}) we have modified this condition by introducing a ${\rm O } \left( \epsilon \right)$ contribution to the mass term. Below, we obtain the value of $\delta_m$ that produces a spectral tilt in agreement with observations in the inflationary background. This is not an additional fine-tuning, since it replaces the fine-tuning constraint  (\ref{alpha-ns}). 

The function $F\left(x\right)$ in (\ref{eom-dc-eps}) encodes the first slow roll correction to the mode functions. The time dependence of $F\left( x \right)$ is only due to the $\frac{m^2}{H^2}$ term in eq. (\ref{dc-eom}). It is due to the fact that the mass term $m$ is constant, while $H$ is decreasing.  Even if  $\delta_m$ in eq.~(\ref{m-alpha-eps}) is tuned to reproduce the observed $n_s$ at the CMB scales, the increase of $\frac{m^2}{H^2\left( t \right)}$ with time causes  the mass term to become progressively more relevant for the modes at progressively smaller scales, causing a running of the spectral tilt at first order in slow roll. 

We can see all this from the explicit solution of eq. (\ref{eom-dc-eps}). We write this solution in the form 
\begin{equation}
\delta \chi_k \left( x \right) \equiv   \left[ 1 + \epsilon \, \beta \left( x \right)  \right]  \, \delta \chi_{k,dS} \left( x \right) 
\,, 
\end{equation}
where $ \delta \chi_{k,dS} \left( x \right) $ is the de Sitter solution (\ref{dc-sol}).  By introducting the primitive 
\begin{equation}
\Xi \left( x \right) \equiv \int \frac{d x}{\delta \chi_{k,dS}^2 \left( x \right) } = i k {\rm e}^{-2 i x} \, \frac{1+i x}{1-i x} \;\;, 
\end{equation} 
it is easy to verify that the solution is given by 
\begin{eqnarray}
\beta \left( x \right) & = &  - \Xi \left( x \right) \int_x^\infty d y \, F \left( y \right)  \delta \chi_{k,dS}^2 \left( y \right) \nonumber\\ 
& & \quad\quad + \int_x^\infty d y \, \Xi \left( y \right) F \left( y \right)  \delta \chi_{k,dS}^2 \left( y \right) \,, 
\label{beta-eps}
\end{eqnarray} 
where the integration constants are chosen so that $\beta(x) = \beta'(x) = 0$ at past infinity.  The solution contributes at ${\rm O } \left( \epsilon \right)$ to the curvaton power spectrum evaluated at the end of inflation
\begin{equation} 
P_{\zeta_\sigma} \propto k^3 \vert \delta \chi_k \vert^2 \propto 1 + 2 \,\epsilon\, {\rm Re }\left[ \, \beta \left( \frac{k}{H} \right)\right] \,, 
\label{Pz-eps}
\end{equation} 
where the proportionality constants are $k-$independent. 

The two integrals in (\ref{beta-eps}) can be straightforwardly computed, but the full expression is lengthy and not particularly illuminating. We are interested only in the sub-horizon limit ($x \rightarrow 0^+$) of this solution, which, once inserted into (\ref{Pz-eps}), gives 
\begin{eqnarray}
n_{s} - 1 & \equiv & \frac{\partial {\rm ln }  P_{\zeta_\sigma}}{\partial {\rm ln } k }    = 
\frac{2}{3} \,\epsilon\Big\{  2 \alpha \left( \alpha + 3 \right) \log \frac{k}{H} 
- F_0 \nonumber\\ 
& & + \frac{\alpha \left( \alpha + 3 \right)}{3} \left( - 14 + 6  \gamma_E + \log 64 \right) 
\Big\} \;,   
\label{ns-eps}
\end{eqnarray}
where $\gamma_E$ is the Euler-Mascheroni constant, and 
\begin{eqnarray}
\alpha_s & \equiv & \frac{\partial n_s}{\partial {\rm ln } k } = 4\,\epsilon\,\left(\alpha+\frac{\alpha^2}{3}\right)=\frac{4\,\epsilon\,m^2}{3\,H^2} \;.   
\label{as-eps}
\end{eqnarray} 

These expressions are accurate at ${\rm O } \left( \epsilon \right)$ (and therefore we can disregard the time variation of $H$ during inflation in them). By inserting the expression (\ref{eom-dc-eps}) for $F_0$ into   (\ref{ns-eps}), we see that the mass  contribution  $\delta_m$ can be chosen so that the spectral tilt  (\ref{ns-eps}) matches the observed result (\ref{ns-as-Planck}) at the given CMB pivot scale.   On the other hand, the running $\alpha_s$ is independent of $\delta_m$ and  ${\rm O } \left( \epsilon\,{m^2}/{H^2} \right)$, hence first order in slow-roll since the ratio $m^2/H^2$ needs not to be small in our scenario.  This implies, as anticipated, that $\alpha_s$ can be much larger than the typical values obtained in inflationary set-ups. Notice that $\alpha_s$ as given in eq.~(\ref{as-eps})  is positive for a positive value of $m^2$.  This implies that the amplitude of the perturbations at the very small scales that left the horizon towards to end of inflation can be much larger than that observed at COBE scales. In particular, if $\epsilon$ or $m^2$ are so large that the amplitude of the power spectrum  in the range $P_\zeta \ge 10^{-2}$ at the end of inflation, then primordial black holes would be overproduced (see for example \cite{Peiris:2008be,Josan:2009qn}). 

Inserting in  the parametrization (\ref{parametrizationP}), the central value  $n_s-1 = -0.0397$ given in 
(\ref{ns-as-Planck}), and the COBE normalization $P_\zeta \left( k_0 \right) \simeq \left( 5 \cdot 10^{-5} \right)^2$, we see that  $P_\zeta = 10^{-2}$ can be reached at the shortest scales produced during inflation (namely for  
$k = {\rm e}^{60} k_0$) if $\alpha_s \ga 0.01$. This can be used as a phenomenological limit on $\alpha_s$, and on the parameters $\frac{m}{H}$ and $\epsilon$ that lead to the result (\ref{as-eps}) obtained in this Section.

Note that the positivity of $\alpha_s$ derives from the assumption that $\alpha$ is constant. In general, the relation~(\ref{N-phi}) can be only approximate, in which case $\alpha$ will acquire a weak time dependence. This in turn would lead to a model dependent modification of the results of the previous section, whose analysis goes behind the scope of the present paper. It is nevertheless worth mentioning that, if we allow the parameter $\alpha$ to acquire a slow-roll suppressed time dependence, then the running $\alpha_s$ will acquire a more general expression, which can be of either sign.

\section{An explicit realization}
 \label{sec:realization}

In this Section we present an explicit model with an action of the form~(\ref{action2}), providing an existence proof of our mechanism that is minimal in matter content. 

We consider a complex scalar $\Phi$ carrying a global $U(1)$ and with a nontrivial kinetic term. We assume that some dynamics breaks the $U(1)$ while preserving a $Z_2$ symmetry. As a consequence the Lagrangian of the system reads
\begin{align}
{\cal L}= f^2\left(\left|\Phi\right|\right)\,&\left[-\partial_\mu\Phi\,\partial^\mu\Phi^*-V\left(\left|\Phi\right|\right)+\frac{\mu^2}{4}\,\left(\Phi-\Phi^*\right)^2\right]\,.
\end{align}
If we decompose the complex field $\Phi$ into a radial and an angular part $\Phi=\rho\,e^{i\theta}$, and we consider $|\theta|\ll 1$ so that it is close to the minimum  of its potential, we obtain a Lagrangian  analogous to that of~(\ref{action2}):
\begin{equation}
{\cal L}= f^2\left(\rho\right)\,\left[-\partial_\mu\rho\,\partial^\mu\rho-\rho^2\left(\partial_\mu\theta\,\partial^\mu\theta+\mu^2\,\theta^2\right)-V\left(\rho\right)\right]\,.
\label{ex1-L}
\end{equation}

Since the mechanism described in the previous sections requires the coefficient of the kinetic term of the curvaton to vary rapidly during inflation, we assume that inflation takes place near some point $\rho=\rho_0$ where the function $f (\rho)$ is singular, and  we approximate 
\begin{equation}
 f (\rho)\simeq \frac{1}{\sqrt{2}}\,\frac{1}{\rho/\rho_0-1}\,,
\end{equation}
where $\rho_0>0$. Then the canonically normalized inflaton reads
\begin{equation}
\phi=\rho_0\,\log(\rho/\rho_0-1)\,,
\end{equation}
where we have chosen the branch $\rho>\rho_0$ to make sure that the kinetic term for $\theta$, that is proportional to $\rho^2$, does not cross zero. In terms of $\phi$, the Lagrangian reads
\begin{align}
{\cal L}=&\frac{\rho_0^2}{2}\,\left(e^{-\phi/\rho_0}+1\right)^2\left[-\partial_\mu\theta\,\partial^\mu\theta-\mu^2\,\theta^2\right]\nonumber\\
&-\frac{1}{2}\partial_\mu\phi\,\partial^\mu\phi-\frac{e^{-2\phi/\rho_0}}{2}\,V\left(\rho_0(1+e^{\phi/\rho_0})\right)\,,
\end{align}
and we will be interested in the region where $\phi/\rho_0$ is large and negative (corresponding to $\rho \rightarrow \rho_0^+$), so that the coefficient of the kinetic term for $\theta$ is rapidly running during inflation. Now, if $V$ is a generic function of $\rho$, the potential will be too steep to support inflation. In order to obtain inflation we choose the (tuned) form
\begin{equation}
V(\rho)=\Lambda^4\left(\frac{\rho}{\rho_0}-1\right)^{2-\delta}\,,
\label{ex1-V}
\end{equation}
with $\delta$ a positive number much smaller than unity. Then the potential in terms of $\phi$ becomes
\begin{equation}
  f^2 \left( \rho \right) V \left( \rho \right) \equiv 
V(\phi)=\frac{\Lambda^4}{2}\,e^{-\phi\,\, \delta/\rho_0}\,.
\label{ex1-V2}
\end{equation}
Namely, we obtain power law inflation in an exponential potential \cite{Lucchin:1984yf},~\footnote{Incidentally we point out that, while power law inflation  \cite{Lucchin:1984yf}, under the assumption that the inflaton is responsible for the observed perturbations,  is now ruled out \cite{Ade:2013uln}, this model is still viable under the curvaton hypothesis that we are making here.}  characterized by the constant slow-roll parameter $\epsilon = \frac{M_p^2 \delta^2}{2 \rho_0^2}$, and by the analytic solution 
\begin{equation}
\phi = \frac{\sqrt{2} M_p}{\sqrt{\epsilon}} \log  \left[ \frac{\epsilon}{\sqrt{3-\epsilon}} \, \frac{\Lambda^2}{\sqrt{2} M_p} t \right] \;\;,\;\; H = \frac{1}{\epsilon \, t} \;\;. 
\label{phi-H-sol}
\end{equation} 
Strictly speaking, inflation never ends in an exact exponential potential,  so we assume that the potential  is modified at some value $\phi_{\rm end} \ll - \rho_0$ (namely, very close to the pole of (\ref{ex1-V})), where inflation ends. In the regime  $\phi \ll - \rho_0  $, the curvaton  Lagrangian    reads 
\begin{eqnarray}
{\cal L}_\sigma & \simeq & \frac{1}{2} {\rm e}^{\frac{2}{\rho_0} \left( \phi_{\rm end} - \phi \right) } \left[ - \partial_\mu \sigma  \partial^\mu \sigma - \mu^2 \sigma^2 \right] \;\;, \nonumber\\
 \sigma & \equiv & {\rm e}^{-\frac{\phi_{\rm end}}{\rho_0}} \, \rho_0 \,  \theta \;\;, 
\label{ex1-Ls}
\end{eqnarray} 
with $\sigma$ normalized in such a way that   ${\cal N} = 1$ at the end of inflation.

Using eq. (\ref{N-phi}) we can see that 
\begin{equation}
\alpha = \frac{ \sqrt{2 \epsilon} M_p }{\rho_0 } \;\;, 
\end{equation} 
and, despite the slow-roll suppression, $\alpha$ can be large (corresponding to large $\mu/H$) for sufficiently large $M_p/\rho_0$. 

To conclude, let us verify two assumptions that  we have imposed to obtain this result. To retain only the quadratic term in $\theta$ in eq. (\ref{ex1-L}), the assumption $\vert \theta \vert \ll 1$ was made. Using eqs. (\ref{phi-chi}), (\ref{chi-0}), and (\ref{ex1-Ls}), we have 
\begin{equation}
\theta \left( t \right) = \frac{C}{\rho_0} \, {\rm e}^\frac{\phi \left(  t \right)}{\rho_0} \;\;. 
\end{equation} 
and, due to the strong exponential suppression (recall that $\phi \ll - \rho_0  $), the condition $\theta \ll 1$ is easily met. Secondly, as we have discussed in Section~\ref{sec:replace}, the condition (\ref{Ds2-gamma}) must be imposed. For the present model, this condition translates into $\frac{\epsilon^{3/2} M_p}{\rho_0} \ll \frac{3 \sqrt{3}}{16 \sqrt{N_{\rm CMB}}}$, which can easily be satisfied even for $\alpha \gg 1$.  

\smallskip

Before concluding this section let us point out that a kinetic term for the curvaton of the form appearing in the action~(\ref{action2}) naturally emerges in models of supergravity with a K\"ahler potential of the form (in units of $M_P=1$)
\begin{equation}
K=-\kappa\,\log(T+T^*)\,,
\end{equation}
with $\kappa$ a positive constant. In this case, in fact, if we denote by $\phi$ the canonically normalized real component of the field $T$, the imaginary part $\sigma$ of $T$ gets a kinetic term $\propto e^{-2\,\sqrt{2/\kappa}\,\phi}\,(\partial\sigma)^2$. Our mechanism will then be at work in situations where a mass term for $\sigma$ is also multiplied by a similar factor $\sim e^{-2\,\sqrt{2/\kappa}\,\phi}$.

Finally, one more construction leading to an action of the form~(\ref{action2}) can be realized in models with extra dimensions. Consider a model in which the size of the extra dimension(s) plays the role of the inflaton $\phi$, while the curvaton $\sigma$ is a bulk field. It is simple to check explicitly  that if the kinetic term for the scalar field $\sigma$ is localized on the brane, whereas the operator controlling its mass $m$  lives in the bulk, then the low energy effective Lagrangian has the form~(\ref{action2}).

\section{A change of  frame}
 \label{sub:frame}

In the previous Sections we considered realizations of a two-field system for which the curvaton Lagrangian is multiplied by a suitable function of the inflaton. In this Section, we take another point of view, and ask whether a suitable coordinate transformation provides a second perspective on the effects of  ${\cal N}$ as a pure function of time -- without interpreting it as function of the inflaton field. We start with action (\ref{action1}), with ${\cal N}$ given in  eq.~(\ref{N-t}), and use the time variable $\omega$, related to the physical time $t$ by  $d\,\omega\,=d\, t/{\cal N}^2(t)$. The resulting action describes a scalar field with time dependent speed of sound and time dependent mass
\begin{equation} 
{\cal S}=\int \,d \omega\, d^3 x\, a^3\,\left(\frac{\dot{\sigma}{}^2}{2}-c^2_s(\omega)\,\frac{(\nabla\sigma)^2}{2}-M^2(\omega)\,\frac{\sigma^2}{2}\right)
\end{equation}
with $\dot{\sigma}\,\equiv\, d\,\sigma/d \omega $ and
\begin{eqnarray} c_s(\omega)&=&a^{2\alpha}(\omega)\label{tdss}\,,\\
M(\omega)&=&a^{2\alpha}(\omega)\,m \label{tdm}\,.
\end{eqnarray}

Let us specify the discussion to de Sitter space, with the scale factor $a(t)\,=\,e^{H t }$. Then the relation between the two time variables is 
\begin{equation} 
t\,=\,\omega_0-\frac{1}{2\,\alpha\,H}\, \ln{\left(1-2\,\alpha\,H\,(\omega-\omega_0)\right)}\,. 
\end{equation} 
with $\omega_0$ an arbitrary integration constant. 
Notice that when $\alpha\to 0$, one correctly finds $t\to\omega$. When choosing $\omega_0=-1/(2\alpha H)$, the scale factor appearing in the metric
$d s^2\,=\, - d \omega ^2+a^2(\omega)\,d \vec{x}^2 $ reads, as a function of $\omega $,
\begin{equation} 
a(\omega)\,=\,\left( -2\,e\,\alpha\,{H} \, \omega\right)^{-\frac{1}{2\alpha}}\,.
\end{equation}
Hence the underlying geometry is no more de Sitter space, but power-law expansion. In this frame, we learn that our results can also be obtained by means of a scalar field embedded in a FRW space-time with power-law expansion, provided that the scalar mass and sound speed vary appropriately with time. This example is related to and generalizes the results of~\cite{Khoury:2008wj}. While in that work a scale invariant spectrum was obtained in a Universe with arbitrary power-law expansion for a degree of freedom with time dependent speed of sound and zero mass (modeled as a fluid with constant equation of state), in this section we have shown that our system is analogous to that of~\cite{Khoury:2008wj} with the addition of a nontrivially evolving mass for the scalar.

\section{Conclusions}
 \label{sec:conclusions}%

The calculation of the mass spectrum of a model of inflation can be misleading when performed on a time-independent background. In the present work we have shown that a field whose mass $m$, when computed on a static background, is comparable to or larger than $\sqrt{V/3\,M_P^2}$, can still acquire a scale invariant spectrum of perturbations, provided its Lagrangian is multiplied by a suitably time-dependent function. This is achieved via the relation~(\ref{alpha-cond}), that represents a generalization of the fine-tuning to very small values of the mass of the curvaton. As we have discussed in Sections~\ref{sec:replace} and \ref{sec:run}, this can be achieved in a self consistent way and without conflicting with observations provided the parameters of the model are such that the fluctuations in the inflaton are much smaller than those in the curvaton, eq.~(\ref{Ds2-gamma}). In Section~\ref{sec:run} we shown that our mechanism leads a distinctive feature that can be tested by future observations, namely a large running of the spectral index $\alpha_s$,  proportional to the first power of slow-roll parameters, hence of a size comparable to  $n_s-1$. Indeed, the leading contribution to this parameter is $\alpha_s \,\propto\,\epsilon\,\, m^2/H^2$, where $\epsilon$ as usual parameterizes the time dependence of the Hubble parameter during inflation. In standard inflationary scenarios, $m^2/H^2$ is small (being associated to the slow-roll parameter $\eta$) while in our case it can be of order one or more, making $\alpha_s$ large but still compatible with current observations that do not exclude  a large value for this quantity. 

A kinetic term with the structure of~(\ref{action2}) appears for instance when the inflaton  $\phi$ measures the volume spanned, in field space, by $\sigma$. This is apparent in the example of Section~\ref{sec:realization}, where the inflaton is related to the radial component and the curvaton to the angular component of a complex field.  Hence, an almost scale invariant spectrum of perturbations and small non-Gaussianity can be a manifestation of particular interactions among the fields producing inflation and generating the density perturbations. This approach might be useful to generalize the question of naturalness of inflationary constructions by allowing a large mass for the fields involved in sourcing primordial curvature fluctuations, provided the coupling between inflaton and curvaton is appropriately tuned. Moreover, such scenario can be tested and possibly  corroborated if future observations will favor higher-than-expected values for the running of the spectral index $\alpha_s$. 

\vskip.35cm
\noindent{\bf Acknowledgements} 

\smallskip

It is a pleasure to thank Rob Crittenden, David Langlois, Sami Nurmi, and David Wands for useful comments.  The work of  M.P. was partially supported by the DOE grant DE-FG02-94ER-40823 at the University of Minnesota. The work of L.S. is partially supported by the U.S. National Science Foundation grant PHY-1205986. G.T. is supported by an STFC Advanced Fellowship ST/H005498/1.

\appendix%
\section{Modification of $\langle \sigma^2 \rangle$ and $\langle \sigma^3 \rangle$ from the inflaton-curvaton coupling}
\label{app:s2s3}

In this Appendix we develop the calculations leading to the results (\ref{s2s3-res}) of the main text. We start from the action (\ref{action2}) and we perform this computation  in the interaction picture, using the in-in formalism \cite{Weinberg:2005vy}. At the unperturbed level, ${\cal N}$ is treated as an external function of time, and the two fields $\phi,\sigma$ are quantized independently. Specifically, we decompose the fields according to 
\begin{eqnarray} 
\sigma(\tau, \vec{x}) & \equiv &   \frac{ 
 \chi^{(0)} \left( \tau \right) + \delta \chi \left( \tau , \vec{x} \right)  }{  a \, \langle {\cal N} \rangle }  \;\;, \nonumber\\ 
\phi \left( \tau , \vec{x} \right) & \equiv & \phi^{(0)} \left( \tau \right) + \frac{v \left( \tau , \vec{x} \right) }{a} \;\;, 
\end{eqnarray}
and we stress that only the background value of ${\cal N}$ is used in the free field decomposition and quantization. Therefore, the quantization of $\sigma$ proceeds as in  Section \ref{sec:PS}, and the result  (\ref{P-zeta}) for the power spectrum holds at zeroth order in the $\phi-\sigma$ interaction. For the  inflaton field, we have  
\begin{align}
& v_F  = \int \frac{d^3 k}{\left( 2 \pi \right)^{3/2} }\, {\rm e}^{i \vec{x} \cdot \vec{k}} \,  {\hat v}_{k,F}\;\;,\;\;\nonumber\\
& {\hat v}_{k,F} = v_{k,F}  \left( \tau \right) {\hat b}_{\vec{k}} +   v_{k,F}^* \left( \tau \right) {\hat b}_{-\vec{k}}^\dagger  \;\;, \nonumber\\ 
& v_{k,F} =  \frac{{\rm e}^{-i k \tau}}{\sqrt{2 k}} \left( i + \frac{1}{k \tau} \right) \;\;, 
\end{align} 
where the  annihilation and creation operators obey the same algebra as those of $\sigma$, and where we have disregarded slow roll corrections in the mode function solution. The suffix $F$ in these expressions indicates that we are quantizing the free field according to the  interaction picture procedure. Accordingly, we use the suffix $F$ for indicating the free field $\delta {\hat \chi}_F$ and free mode function $\delta \chi_F$ entering in the quantization of $\sigma$. We note that $v_{k,F} = \delta \chi_{k,F}$, because, due to the tuning (\ref{alpha-cond}), $\chi$ effectively behaves as a massless field, and the same is true for the inflaton at zeroth order in slow roll. 

According to the in-in formalism, the correlators of $\delta {\hat \chi}$ are related to that of the free field $\delta {\hat \chi}_F$ by 
\begin{eqnarray}
& & \left \langle \delta {\hat \chi}_{\vec{k}_1}  \dots  \delta {\hat \chi}_{\vec{k}_n} \left( \tau \right) \right \rangle  =  \sum_{N=0}^{\rm \infty} \left( - i \right)^N \int^\tau d \tau_1 \dots \int^{\tau_{N-1}} d \tau_N \nonumber\\ 
& & \left \langle \left[ \left[ \dots \left[ 
  \delta {\hat \chi}_{F, \vec{k}_1}  \dots  \delta {\hat \chi}_{F,\vec{k}_n} \left( \tau  \right) 
  , H_{\rm int } \left( \tau_1 \right) \right] , \dots \right] , H_{\rm int} \left( \tau_N \right) \right] \right \rangle \,, \nonumber\\ 
\label{in-in}
\end{eqnarray} 
where $H_{\rm int} = - \int d^3 x\, {\cal L}_{\rm int}$. The quantity ${\cal L}_{\rm int}$ is the interaction Lagrangian of the system, which is the difference between the Lagrangian in~(\ref{action2}) and the free Lagrangian. In turn, the free Lagrangian is given by the Lagrangian in~(\ref{action2}) with ${\cal N}$ replaced by $\langle {\cal N} \rangle$.~\footnote{With this last statement, we are disregarding the inflaton self-interaction, which provide subleading corrections to  $\delta \chi$ correlators. In our computations we are also disregarding the mixing of $\delta \chi$ and $\delta \phi$ with the metric perturbations. These interactions are of gravitational strength, and are known to lead to unobservably small non-Gaussianity. As we now see, the inflaton-curvaton interactions are enhanced by inverse powers of the slow roll parameter $\epsilon$ with respect to gravitational interactions, and for this reason metric perturbations can be consistently disregarded. The situation is analogous to that studied in \cite{Bartolo:2012sd}, where an analogous interaction is present between the inflaton and a vector field.} 

As usual in perturbation theory, we expect that the strongest corrections to any given correlator are provided by terms with the fewest number of fields in the right hand side of eq. (\ref{in-in}).  The lowest order interaction  between the two fields is the quadratic term 
\begin{equation}
H_{\rm int} \supset - \frac{4}{3} \, \frac{\alpha^2 \gamma}{\tau^2} \, \int d^3 p \; {\hat \Pi}_{F,\vec{p}} \; {\hat v}_{F,-\vec{p}} \;\;, 
\label{mass-ins}
\end{equation}
where, to shorten the notation, we have defined the auxiliary field 
\begin{equation}
\Pi \equiv \frac{\delta  \chi'}{a H} + 2 \delta  \chi \;\;. 
\end{equation}
The interaction (\ref{mass-ins}) corresponds to a mass insertion, and it is proportional to the curvaton {\it vev} $C \propto \gamma$ (see eq.~(\ref{gamma})). The first correction to the two point function of $\delta \chi$ is obtained by taking two such interactions terms in (\ref{in-in}). This corresponds a Feynman diagram with an  external $\delta \chi$ mode, which is `converted' into a $v$ mode by a mass insertion, which is then converted back to a $\delta \chi$ mode by a second mass insertion. The corresponding correction is 
\begin{eqnarray} 
& & \Delta \left\langle \delta {\hat \chi}_{\vec{k}_1}  \delta {\hat \chi}_{\vec{k}_2} \left( \tau \right) \right\rangle  =  
\frac{16 \alpha^4 C^2}{\epsilon M_p^2} \delta^{(3)} \left( \vec{k}_1 + \vec{k}_2 \right) \nonumber\\ 
& & \quad \times \int_{ \tau_{\rm in}}^\tau \frac{d \tau_1}{\tau_1^2} 
 {\rm Im } \left[ \delta \chi_{F,k_1} \left( \tau \right) \Pi_{F,k_1}^* \left( \tau_1 \right) \right] 
 \\ 
& & \quad \times   \int_{\tau_{\rm in}}^{\tau_1} \frac{d \tau_2}{\tau_2^2}  
{\rm Im } \left[  \delta \chi_{F,k_1} \left( \tau \right)  \delta \chi_{F,k_1} \left( \tau_1 \right) 
 \delta \chi_{F,k_1}^* \left( \tau_2 \right)   \Pi_{F,k_1}^* \left( \tau_2 \right) \right] \,, \nonumber 
\label{DP-par}
\end{eqnarray} 
where  $\tau_{\rm in}$ is some initial time during inflation at which the mode was deeply inside the horizon  (namely, $\vert k_1 \tau_{\rm in} \vert \gg 1$). 

We evaluated (\ref{DP-par}) in two different ways. Firstly, we performed the  $\tau_2$ integration analytically, and the $\tau_1$ integration numerically. In this last step, one notices that the integral is dominated by the latest times, in which the mode is well outside the horizon, while it is converging in the UV ($\tau_1 \rightarrow \tau_{\rm in}$). Justified by this observation, we expanded the full integrand of (\ref{DP-par}) in the super horizon regime ($\vert k_1 \tau \vert \; , \; \vert k_1 \tau_i \vert \ll 1$), and retained only the leading term. In this way both integrals can be trivially performed analytically. Both computations lead to 
\begin{equation}
\Delta   \left\langle \delta {\hat \chi}_{\vec{k}_1}  \delta {\hat \chi}_{\vec{k}_2} \left( \tau \right) \right\rangle  \simeq  
- \frac{16\, \alpha^4 H^2 \gamma^2}{27  } \, \frac{\delta^{(3)} \left( \vec{k}_1 + \vec{k}_2 \right)}{k_1^3} 
a^2 N_{\rm CMB}\,,   
\label{Dchi2}
\end{equation} 
where $N_{\rm CMB} \equiv {\rm ln } \frac{1}{-k_1 \, \tau}$ is the number of e-folds before the end of inflation when the mode $k_1$ left the horizon. Adding this to the free field result, 
\begin{equation}  
 \left\langle \delta {\hat \chi}_{F,\vec{k}_1}  \delta {\hat \chi}_{F,\vec{k}_2} \left( \tau \right) \right\rangle  =  
\frac{H^2}{2} a^2 \left( \tau \right)  \, \frac{\delta^{(3)} \left( \vec{k}_1 + \vec{k}_2 \right)}{k_1^3} \, , 
\label{chi2-free}
\end{equation}
and using eq.~(\ref{zetasig}), we obtain the result for $\langle \zeta_\sigma^2 \rangle$ in eq.~(\ref{s2s3-res}) of the main text.  

Next, we want to compute the three-point correlation function. To obtain the dominant contribution, we need to expand the interaction hamiltonian to one higher order in the fields. We obtain the term 
\begin{eqnarray} 
H_{\rm int} & \supset & 
- \frac{ \alpha H}{M_p \sqrt{2 \epsilon} \tau } \int \frac{d^3 p d^3 q}{\left( 2 \pi \right)^{3/2}} 
\; {\hat v}_{F,-\vec{p}-\vec{q}} 
\Bigg\{ {\hat \Pi}_{F,\vec{p}} {\hat \Pi}_{F,\vec{q}} \nonumber\\
 & & \!\!\!\!\!\!\!\! \!\!\!\!\!\!\!\! 
  - 2 \left( 3 + \alpha \right) {\hat \Pi}_{F,\vec{p}}  \delta {\hat \chi}_{F,\vec{q}}  
+ \left[ \tau^2 \vec{p} \cdot \vec{q}   + 3 \left( 3 + \alpha \right)    \right]  \delta {\hat \chi}_{F,\vec{p}}   \delta {\hat \chi}_{F,\vec{q}}    \Bigg\}   \,,  \nonumber\\  
\end{eqnarray} 
which, combined with the mass insertion (\ref{mass-ins}), gives the contribution in the second diagram of Figure~\ref{fig:diagrams}.~\footnote{We note that  a ${\rm O } \left( \delta \chi \, v^2 \right)$ term is also present in $H_{\rm int}$, which can be combined with two mass insertions to give a different tree level contribution to $\delta \chi^3$. We expect that - as long as the correction (\ref{Dchi2}) is subdominant to the free field result (\ref{chi2-free}) - this contribution is suppressed with respect to the one we are computing, as it involves more mass insertions.} This evaluates to 
\begin{eqnarray} 
\left\langle \delta {\hat \chi}_{\vec{k}_2} \delta {\hat \chi}_{\vec{k}_2} \delta {\hat \chi}_{\vec{k}_3}  \right\rangle & \simeq & - \frac{\alpha^4 H \gamma}{6 \pi^{3/2} \sqrt{\epsilon} M_p} \, \frac{\sum_i k_i^3}{\prod_i k_i^3} \delta^{(3)} \left( \vec{k}_1 + \vec{k}_2 + \vec{k}_3 \right) \nonumber\\ 
& & \quad\quad  \int^\tau \frac{d \tau_1}{\tau_1^4}  \int^{\tau_1} \frac{d \tau_2}{\tau_2^4} \left( 4 \tau^3 + 3 \tau_1^3 - 3 \tau_2^3 \right)  \nonumber\\ 
& & \!\!\!\!\!\!\!\!  \!\!\!\!\!\!\!\! 
\simeq   \frac{\alpha^4 H \gamma}{6 \pi^{3/2} \sqrt{\epsilon} M_p} N_{\rm CMB} \, \frac{\sum_i k_i^3}{\prod_i k_i^3} \delta^{(3)} \left( \vec{k}_1 + \vec{k}_2 + \vec{k}_3 \right) \,, \nonumber\\ 
\label{chi3}
\end{eqnarray} 
where, based on what we found for the two-point function, we have restricted the integration to the super-horizon regime, and where for simplicity we have disregarded the difference between the various momenta in the logarithmic factor $N_{\rm CMB} \simeq -\frac{1}{k_i \tau}$.~\footnote{Both  (\ref{Dchi2}) and (\ref{chi3}) neglect ${\rm O } \left( 1 \right)$ factors in comparison with the logarithmic enhancement $N_{\rm CMB}$.} From this expression we obtain the  result for $\langle \zeta_\sigma^3 \rangle$ in eq.  (\ref{s2s3-res}) of the main text.  

\smallskip

Hence, we learn that corrections the 2 and 3-pt functions for $\delta \chi$ calculated in  the free theory   are proportional to the number of e-folds of inflation since the relevant modes left the horizon, and on coefficients controlling the amplitude of mass insertion  and vertex interactions involving curvaton and inflaton fluctuations (as represented in Fig \ref{fig:diagrams}). These corrections are enhanced by powers of $\alpha$ leading to constraints on the model parameters, as discussed in the main text. The same analysis can be carried on to higher point functions with no conceptual changes: the leading contribution to the 4-pt function, for example, does not involve any mass insertion, hence we expect that it  does not lead to more stringent constraints than the ones discussed in eq. (\ref{Ds2-gamma}).

\end{document}